\documentclass[11pt]{article}
\usepackage{geometry}                
\usepackage{graphicx}
\usepackage{amssymb}
\newcommand{\be}{\begin{equation}}
\newcommand{\ee}{\end{equation}}
\newcommand{\bea}{\begin{eqnarray}}
\newcommand{\eea}{\end{eqnarray}}
\newcommand{\bi}{\begin{itemize}}
\newcommand{\ei}{\end{itemize}}
\def\vtr#1{{\bf #1}}

\def\gsim{\, \rlap{$>$}{\lower 1.1ex\hbox{$\sim$}}\,}
\def\lsim{\, \rlap{$<$}{\lower 1.1ex\hbox{$\sim$}}\,}

\begin{document}


\begin{titlepage}
\hfill preprint : NSF-KITP-07-38\vspace{2cm}
\bigskip
\bigskip
\centerline{\Large \bf  Periodic Gravitational Waves From Small Cosmic String Loops}
\vspace{1cm}

\centerline{\large Florian Dubath}
\bigskip
\centerline{\em Kavli Institute for Theoretical Physics}
\centerline{\em University of California}
\centerline{\em Santa Barbara, CA 93106-4030} \centerline{\tt dubath@kitp.ucsb.edu}
\bigskip
\bigskip
\centerline{\large Jorge V. Rocha}
\bigskip
\centerline{\em Department of Physics}
\centerline{\em University of California}
\centerline{\em Santa Barbara, CA 93106} \centerline{\tt jrocha@physics.ucsb.edu}
\bigskip
\bigskip\bigskip
\bigskip\bigskip


\begin{abstract}
We consider a population of small, high-velocity cosmic string loops. We assume the typical length of these loops is determined by the gravitational radiation scale and use the results of~\cite{Polchinski:2007rg} which pointed out their highly relativistic nature. A study of the gravitational wave emission from such a population is carried out. The large Lorentz boost involved causes the lowest harmonics of the loops to fall within the frequency band of the LIGO detector. Due to this feature the gravitational waves emitted by such loops can be detected in a periodic search rather than in burst or stochastic analysis. 

It is shown that, for interesting values of the string tension ($10^{-10}\lsim G\mu\lsim 10^{-8}$) the detector can observe loops at reasonably high redshifts and that detection is, in principle, possible. We compute the number of expected observations produced by such a process. For a 10 hour search we find that this number is of order $O(10^{-4})$. This is a consequence of the low effective number density of the loops traveling along the line of sight. However, small probabilities of reconnection and longer observation times can improve the result.
\end{abstract}
\end{titlepage}
\baselineskip = 16pt

\section{Introduction}

The existence of  cosmic strings in our Universe is a possibility which has been recently rejuvenated by proposals from string theory~\cite{Sarangi:2002yt,Copeland:2003bj,Polchinski:2004ia,Davis:2005dd}. Such strings form a network which is expected to be in a scaling state with a few dozen long strings and numerous loops~\cite{Albrecht:1989mk,Bennett:1989yp,Allen:1990tv,Hindmarsh:1994re,VilenkinShellard}.  As discussed in~\cite{Polchinski:2007rg}, there is evidence supporting the existence of two distinct populations of loops in a scaling cosmic string network~\footnote{However, see~\cite{Ringeval:2005kr} for a different view.}.  One of the peaks in the distribution consists of large loops with a size just an order of magnitude below the horizon scale as shown in recent numerical simulations~\cite{Vanchurin:2005pa,Olum:2006ix}.  These form occasionally by self-intersection along widely separated points on a string.  In contrast, there is also a population of small loops at the gravitational radiation scale which, in an expanding universe, seems to be more numerous than the former. 
In~\cite{Polchinski:2006ee} the production rate of these tiny loops was found to be divergent and so the proposal is that gravitational radiation acts as a UV cutoff.  It is the small but non-trivial fractal dimension present in the short distance structure that feeds the small loop production.  This process occurs mainly in cusp regions~\footnote{Recall that cusps arise when the left- and right-moving unit vectors ${\bf p}_+(u)$ and ${\bf p}_-(v)$ cross and that (if the reconnection probability is 1) a loop is formed whenever $\int_u^{u+l}du'{\bf p}_+(u') = \int_{v-l}^{v}dv'{\bf p}_-(v')$.} which are therefore excised from the large loops. Due to their production mechanism, the tiny loops are generally expected to feature cusps and kinks, while moving with large Lorentz boosts.

Furthermore, it has been argued that one of the most promising ways to test the presence of a cosmic string network is through its gravitational wave (GW) emission~\cite{Vachaspati:1984gt,Damour:2004kw,Siemens:2006vk,Hogan:2006we}.  One can expect a GW background from the network~\cite{Economou:1991bc,Battye:1997ji,Siemens:2006yp} and the loops may form cusps which emit strong bursts of GW~\cite{Damour:2000wa,Damour:2001bk}. 
The presence of a population of highly boosted loops leads to a peculiar GW signature and therefore deserves a specific study as these may be potential sources for present and planned gravitational wave detectors. Such an investigation is the purpose of this article.

All cosmic string loops are expected to have (pseudo-)periodic behavior, but one of the key features of small high-velocity loops is the fact that their observed period may be as short as a few mili-seconds.  One may even expect that the first few harmonics of the loop enter the frequency band of the GW interferometers.  Such a case is very promising since, on one hand, the lowest harmonics emit the strongest signal and furthermore it has little dependence on the exact loop trajectory: we can expect more robust wave-forms than those originated from the single cusp case (whose behavior under back-reaction or in the presence of small-scale structure has been questioned~\cite{Thompson:1988yj,Quashnock:1990wv,Siemens:2001dx,Siemens:2003ra,Chialva:2006ak}). On the other hand, if this frequency is short enough in order to enter the GW interferometer band, this allows to search for periodic GW signals accumulating a large number of loop periods. As a result the strain sensitivity will be improved by a huge factor corresponding to the square root of the ratio between the observation time and the lifetime of a single cusp.  In addition, the fraction of the small loop population moving directly toward us sees its GW strength boosted.  This leads to a signal-to-noise ratio allowing a direct detection by current detectors for a range of string tension $10^{-10}\lsim G\mu\lsim 10^{-8}$ which is not yet constrained by other experiments and has some promising connections with ${\rm K\!\!\!\!KLM\!\!\!\!\!MT}$ models~\cite{Copeland:2003bj}.

However, the fact that a signal is potentially strong enough is not sufficient to guaranty observations; we need the sources to be numerous enough in order to get a reasonable probability for detection.  Unfortunately, only a small fraction of the observable loops will be moving in a direction close enough to the line of sight to produce a detectable signal.  Thus, we shall see that this probability is too small to expect detection on a regular basis or to efficiently set constraints on the string tension.

The remainder of this article is structured as follows.  In Section~\ref{S_PL} we discuss the properties of small loops and GW emission by them.  In Section~\ref{S_V} we determine under which conditions it is possible to observe such loops in the different LIGO stages.  Finally, we compute the loop density and the probability of detection in Section~\ref{S_LDPD} and then draw some conclusions.

\section{Properties of small loops\label{S_PL}}

Small cosmic string loops move with relativistic center-of-mass velocities~\cite{Polchinski:2007rg} and the large Lorentz factor has some consequences on the analysis of GW detection which were not previously considered in~\cite{Damour:2004kw,Siemens:2006vk,Hogan:2006we,Siemens:2006yp,Garfinkle:1987yw}. 

When considering small scales compared to the Hubble time $t$ one can make the approximation of a flat spacetime and thus represent the string embedding, ${\bf z}(t,\sigma)$, by two functions ${\bf p}_\pm = \partial_t {\bf z} \pm \partial_\sigma {\bf z}$~\cite{VilenkinShellard}.  Adopting the transverse gauge where $\partial_t {\bf z} \cdot \partial_\sigma {\bf z} = 0$, the quantities ${\bf p}_\pm$ are then left- and right-moving unit vectors (which depend only on the combination $u=t+\sigma$ and $v=t-\sigma$, respectively).  To study the effect of the large boost on the GW we will consider a simple model loop, where the functions ${\bf p}_\pm$ are short straight arcs on the unit sphere:
\bea
p_+^\mu(u) & = & \left(1,\sin(Vu),0,\cos(Vu)\right) \simeq \left(1,Vu,0,1-V^2u^2/2\right)\ , \ \ -\beta<u<\beta \nonumber \\
p_-^\mu(v) & = & \left(1,0,\sin(Vv),\cos(Vv)\right) \simeq \left(1,0,Vv,1-V^2v^2/2\right)\ , \ \ -\beta<v<\beta \ .
\label{modelloop}
\eea
These functions are extended to the whole $(u,v)$-plane by imposing periodicity equal to the length of the loop, $2\beta$.  The velocity of the cusp, $V$, and the period of the small loops, $\beta$, were determined in~\cite{Polchinski:2007rg} as average values over the ensemble of small loops.  For a matter-dominated era these quantities are given by
\bea
V & = & 0.08 (G\mu)^{-1.12}t^{-1} \ , \\
\beta & = & 10 (G\mu)^{1.5}t \ ,
\eea
where $t$ is the FRW time.

The embedding of the loop worldsheet in target space is given by
\be
z^\mu(u,v) = \frac12\int p_+^\mu(u)du+\frac12\int p_-^\mu(v)dv \ ,
\label{x/p}
\ee
and is therefore aperiodic as a function of $u$ or $v$ separately:
\bea
z^\mu(u+2\beta,v) & = & z^\mu(u,v+2\beta) = z^\mu(u,v) + \left(\beta,0,0,\sin(V\beta)/V\right) \nonumber \\
& \approx & z^\mu(u,v) + \beta \left(1,0,0,1-V^2\beta^2/6\right).
\label{periodicity}
\eea
However, from the above equation it follows immediately that $z^\mu$ is indeed periodic in $\sigma$, with periodicity $2\beta$.

\begin{figure}
\begin{center}
\includegraphics[width=6cm]{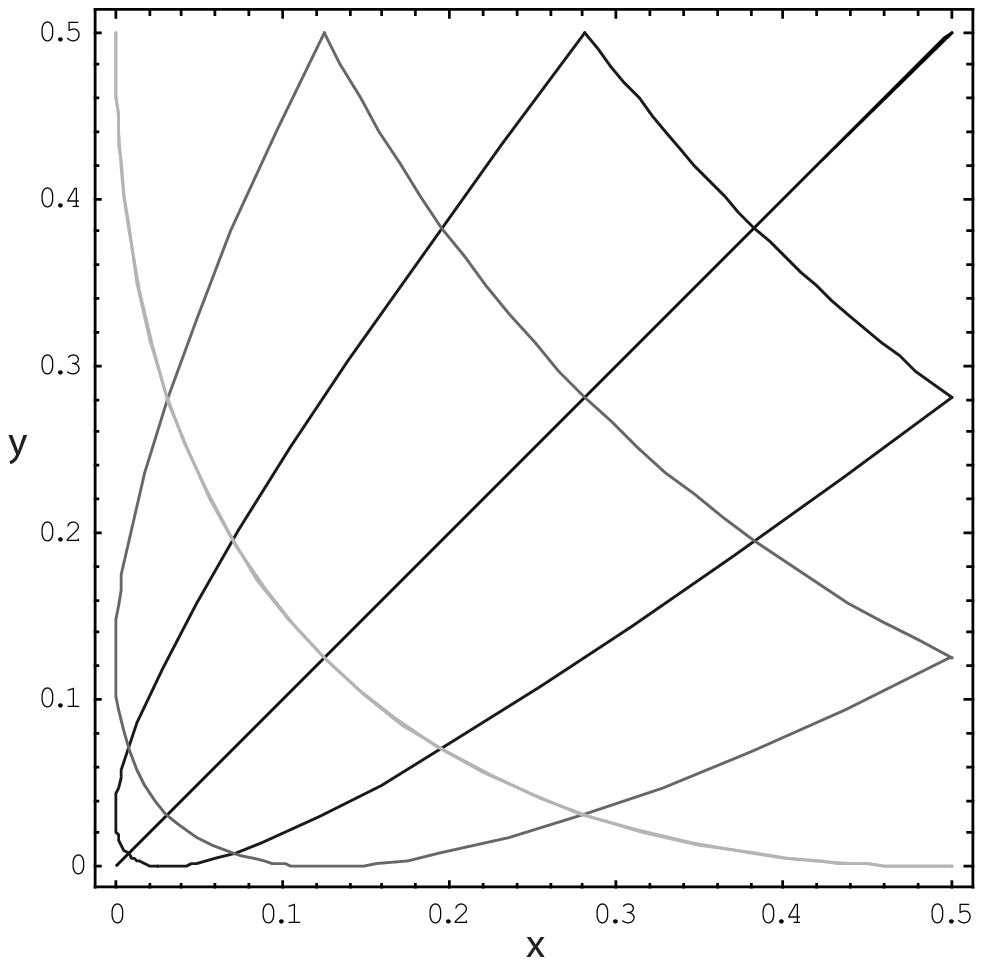}~\includegraphics[width=7cm]{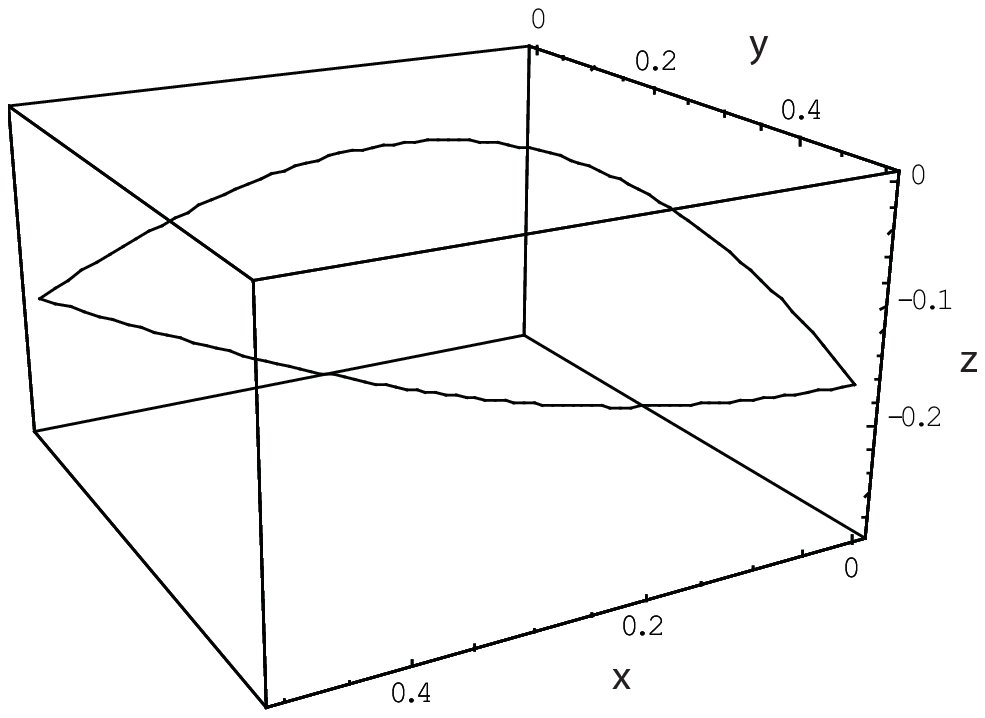}\\
\caption{\label{loop}\footnotesize{The left panel shows the $xy$-projection of four superposed snapshots of the model loop considered.  For clarity a total translation by $(1,1)$ was performed and the $x$- and $y$-axis are in units of $(V\beta)^2$.  The right panel gives a 3D view of the model loop at one of the two moments (the lightest curve in the left panel) at which the $xy$-projection is degenerate.  The $z$-axis was translated by $-V\beta$ and is in units of $(V\beta)^3$.}}
\end{center}
\end{figure}

One can use equations~(\ref{modelloop}) and~(\ref{x/p}) to obtain snapshots of the loop as it evolves in time.  For a fixed time $t_*$, considered to belong to the interval $\left[0,\beta\right]$ for concreteness, the $v$ coordinate may be written in terms of the $u$ coordinate as $v(u) = 2t_* - u$ and then the string is parametrized just by $u$.  If $u \in \left[2t_*-\beta,\beta \right]$,
\be
{\bf z}(u) = \frac{1}{2V} \left( -\cos(Vu) , -\cos(Vv(u)) , \sin(Vu) + \sin(Vv(u)) \right)  \ .
\ee
On the other hand, if $u \in \left[\beta,2t_*+\beta \right]$ we define $u'(u) = u-2\beta \in \left[-\beta,2t_*-\beta \right]$ so that
\be
{\bf z}(u) = \frac{1}{2V} \left( -\cos(Vu'(u)) , -\cos(Vv(u)) , \sin(Vu'(u)) + \sin(Vv(u)) + 2\sin(V\beta) \right)  \ .
\ee
Figure~\ref{loop} shows several snapshots of the model loop~(\ref{modelloop}).  Strictly speaking, there is no self-intersection but at $t=0$ (or equivalently when $u+v=0$) the string actually folds back on itself.  The discontinuities of ${\bf p}_\pm$ at odd multiples of $\beta$ correspond to kinks on the loop traveling in opposite directions.  When they meet the loop becomes degenerate and at the same time a cusp develops at the other end.  However, adding a small (quadratic) perturbation generically renders it non-self-intersecting while the cusp is preserved.  Thus, small cosmic loops are expected to be generally stable against fragmentation.

Small loops with sizes of order $2\beta$ have very large Lorentz factors.  To show this, consider the impulsion of the loop, given by 
\be
P^\mu(u,v) = \frac12\left(p_+^\mu(u)+p_-^\mu(v)\right) \ ,
\ee
and the center-of-mass velocity, which is obtained by averaging the spatial part of $P^\mu(u,v)$ over the worldsheet coordinates,
\be
{\bf v} \simeq \int_0^\beta \frac{dt}{\beta} \int_{0}^{2\beta} \frac{d\sigma}{2\beta} \dot{\bf z}(t,\sigma) 
= \int_{-\beta}^\beta \int_{-\beta}^\beta \frac{du dv}{(2\beta)^2} {\bf P}(u,v) 
= \left(0,0,1-\frac{(V\beta)^2}{6} \right) \ .
\ee
Therefore the boost factor is given by
\be
\gamma=\frac{\sqrt3}{V\beta} \ .
\label{boost}
\ee
For $G\mu = 10^{-9}$ the Lorentz factor~(\ref{boost}) is of order $10^{3}$.  To see the importance of this consider the characteristic frequencies of these loops.  In its rest frame the $n$th harmonic will have frequency $f^{\rm rest}_n=\gamma\beta^{-1}n$, since $\beta$ is its time periodicity in the FRW frame.  Therefore, the observed frequency for a loop moving at an angle $\theta$ with respect to the line of sight is
\be
f_n = \frac{f^{\rm rest}_n}{\gamma\left(1-|\vtr{v}|\cos\theta \right)} \simeq \frac{n}{\beta\left(1-\left(1-\frac{V^2\beta^2}{6}\right)\cos\theta \right)} \ .
\label{harmonics}
\ee
In the special case of a loop whose motion is exactly aligned with the line of sight we see that the observed frequency is boosted by a factor proportional to $\gamma^2$.  This effect can bring the lowest harmonics into the LIGO frequency band for $G\mu\lsim 10^{-9}$, i.e. values of the string tension not yet ruled out by observations (see~\cite{Pogosian:2006hg,Damour:2004kw,Jeong:2006pi,Vanchurin:2005pa} for various bounds), and so the potential for detection arises.

\subsection{Emission of GW}

We now compute the GW spectrum emitted by a loop.  A cosmic string acts as a source term for the gravitational field through its energy-momentum tensor.  For a classical string it is given by~\cite{VilenkinShellard}
\be
T^{\mu\nu}(\vtr{x},t)=\mu\int\left(p^\mu_+p^\nu_-+p^\mu_-p^\nu_+\right)\delta^4\left(x-z(u,v)\right)dudv \ ,
\ee
with $x=(t,\vtr{x})$, while the quadri-vector $z(u,v)$ gives the spacetime location of the world-sheet point with coordinates $(u,v)$.  Since the loop has $2\beta$-periodicity we choose the world-sheet to be a strip $(u,v) \in (-\infty,\infty)\times[-\beta,\beta]$.

Furthermore, for a source of size $\sim d$ localized around the origin, the trace-reversed metric perturbation in the local wave zone ($r\equiv|\vtr{x}|>>d$ ) is given, in the time domain, by~\cite{VilenkinShellard,Misner:1974qy,Weinberg}
\bea
\bar{h}^{\mu\nu}(t,\vtr{x})&=&4G\int \frac{1}{\vert \vtr{x}-\vtr{x}'\vert}T^{\mu\nu}(t-\vert \vtr{x}-\vtr{x}'\vert,\vtr{x}')\, d^3\vtr{x}' \nonumber \\
&\simeq&\frac{4G}{r}\int T^{\mu\nu}(t-r+\vtr{x}'\cdot\vtr{n},\vtr{x}')\, d^3\vtr{x}' \nonumber \\
&=&\frac{4G\mu}{r}\int\left({p}^\mu_+{p}^\nu_-+{p}^\mu_-{p}^\nu_+\right)\delta\left(t-r+\vtr{z}(u,v)\cdot\vtr{n}-z^0(u,v) \right)dudv \ ,
\eea
where $\vtr{n} \equiv \vtr{x}/r$.  When analyzing high frequencies originated by cusps and kinks one must be careful with the fact that the dominant contributions to $p^\mu_\pm$ in the series expansion is a pure gauge term, as was first noted in~\cite{Damour:2001bk}.  However, converting to transverse traceless (TT) gauge eliminates this term\footnote{One can also eliminate the gauge term explicitly by replacing $p^\mu_\pm$ by  $\check{p}^\mu_\pm =p^\mu_\pm-(1,-\vtr{n})$. }.  In any case, we will be interested in the lowest harmonics since these are the frequencies which contribute the most to the observation rate.
Performing a temporal Fourier transform one obtains
\be
\tilde{\bar{h}}^{\mu\nu}(\omega,\vtr{x}) = \frac{4G\mu}{r}e^{i\omega r}\int \left({p}^\mu_+{p}^\nu_-+{p}^\mu_-{p}^\nu_+\right)e^{i\omega (\frac{u+v}{2}-\vtr{z}(u,v)\cdot\vtr{n})} dudv \ .
\ee

Now, decomposing the integral over $u$ into 
\be
\int_{-\infty}^{\infty}du=\sum_{m\in Z\!\!\!Z}\int_{(2m-1)\beta}^{(2m+1)\beta}du \ ,
\ee
using property~(\ref{periodicity}) and defining $\theta$ as the angle subtended between $\vtr{n}$ and the $z$-axis (along which the loop is traveling) we get
\bea
\tilde{\bar{h}}^{\mu\nu}(\omega,\vtr{x})&=&\frac{4G\mu}{r}e^{i\omega r}\sum_{m\in Z\!\!\!Z}e^{i\omega m\beta\left(1-|\vtr{v}|\cos\theta \right)  }    
\underbrace{
\int_{-\beta}^\beta\int_{-\beta}^\beta \left({p}^\mu_+{p}^\nu_-+{p}^\mu_-{p}^\nu_+\right)e^{i\omega (\frac{u+v}{2}-\vtr{z}(u,v)\cdot\vtr{n})} dudv}_{\mathcal{I}^{\mu\nu}(\theta,\phi)} \nonumber \\
& = & \frac{4G\mu}{r}e^{i\omega r}\frac{2\pi}{\beta\left(1-|\vtr{v}|\cos\theta  \right)}\sum_n\delta\left(\omega-\frac{2\pi n}{\beta\left(1-|\vtr{v}|\cos\theta  \right)}\right)\mathcal{I}^{\mu\nu}(\theta,\phi) \nonumber \\
& = & \frac{8\pi G\mu}{r}e^{i\omega r} f_1(\theta)\sum_n\delta\left(\omega-2\pi f_n(\theta)\right)\mathcal{I}^{\mu\nu}(\theta,\phi)\label{h01} \ .
\label{htilde}
\eea
As a check, note that we have recovered the discrete set of frequencies obtained in~(\ref{harmonics}). The computation of $\mathcal{I}^{\mu\nu}(u,v)$ involves the following integrals:
\bea
I^n_{u,j}
&=&\int_{-\beta}^\beta u^je^{2\pi i f_n(\theta)\left(\frac{u}{2}\left(1-\cos\theta\right)-\sin\theta\cos\phi\frac{Vu^2}{4}+\cos\theta\frac{V^2u^3}{12}        \right)}    du\ , \\
I^n_{v,j}
&=&\int_{-\beta}^\beta v^je^{2\pi i f_n(\theta)\left(\frac{v}{2}\left(1-\cos\theta\right)-\sin\theta\sin\phi\frac{Vv^2}{4}+\cos\theta\frac{V^2v^3}{12}        \right)}    dv\ \ \ ,\ \ \  j=0,1,2 \ .
\eea

Before we move on to computing~(\ref{htilde}) one more consideration is in order.  The interaction between a GW and a detector is usually described in the TT gauge and it is convenient to rotate the coordinate frame in order to match the observer description in which the GW arrives along the $z$-axis, so that $\hat{z}^{\rm new}\equiv\vtr{n}$. The rotation matrix that converts between the source frame and the observer frame is given by
\be
R=\left(\begin{array}{ccc}\cos\theta \cos\phi & \cos\theta \sin\phi & -\sin\theta \\
-\sin\phi & \cos\phi & 0 \\
\sin\theta \cos\phi & \sin\theta \sin\phi & \cos\theta \end{array}\right) \ .
\ee
Rewriting $\tilde{\bar{h}}_{\mu\nu}$ in the TT gauge is performed by means of the projector $\Lambda_{ij,kl}$ defined by (see section 10.4.15 in~\cite{Weinberg})
\be
\Lambda_{ij,kl}=P_{ik}P_{jl}-\frac12P_{ij}P_{kl} \ ,
\ee
where $P_{ij}=\delta_{ij}-n_in_j$. Then,
\be
\tilde{h}_{ij}^{TT}=\Lambda_{ij,kl}R_{kk'}R_{ll'}\tilde{\bar{h}}_{k'l'}=\left(\begin{array}{ccc}\tilde{h}_+ & \tilde{h}_\times & 0 \\\tilde{h}_\times & -\tilde{h}_+ & 0 \\0 & 0 & 0\end{array}\right) \ ,
\ee
and we can define the observed strength as
\be
\tilde{h}=\sqrt{|\tilde{h}_+|^2+|\tilde{h}_\times|^2} \ .
\ee

\begin{figure}
\begin{center}
\includegraphics[width=8cm]{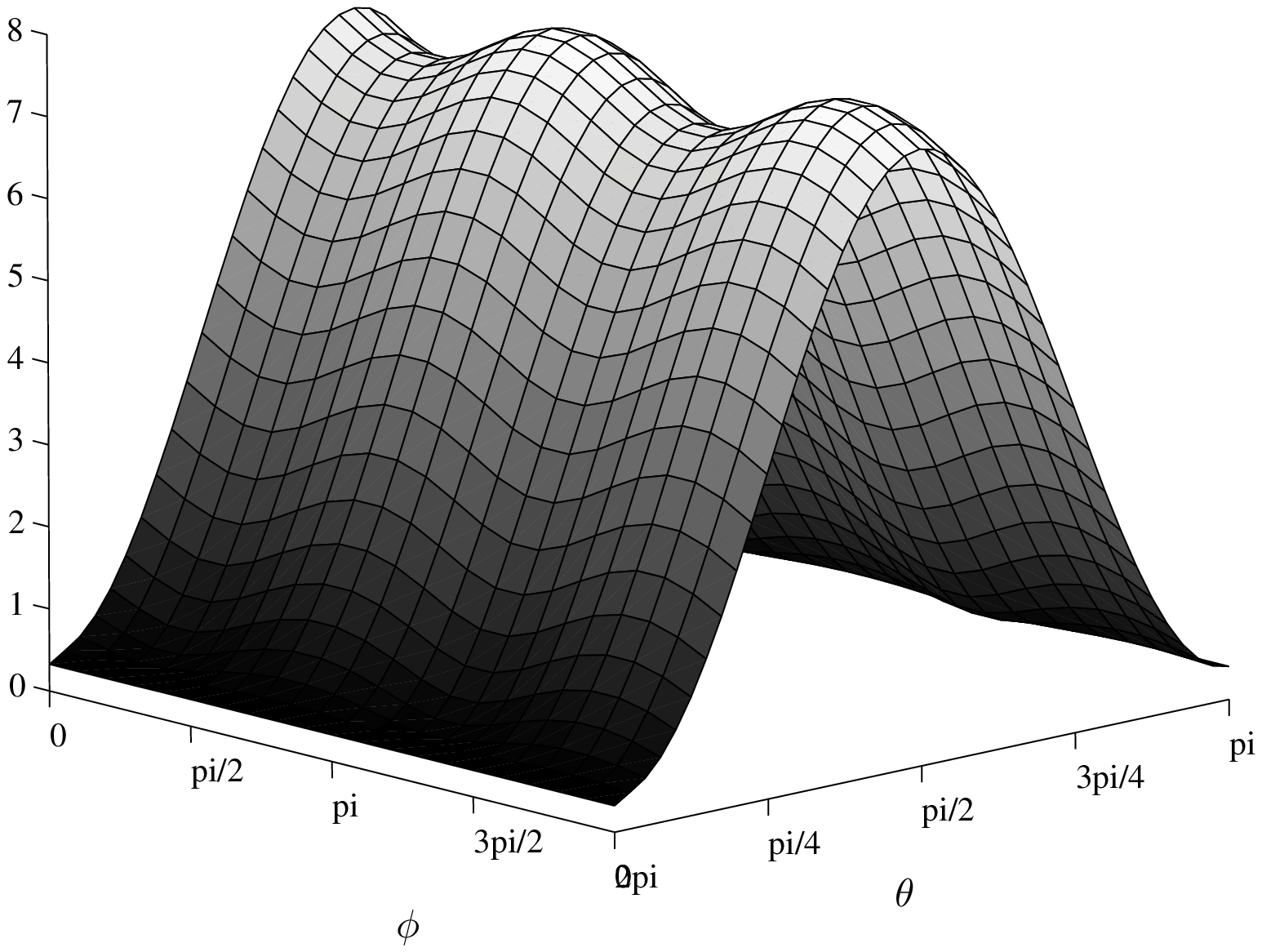}~\includegraphics[width=8cm]{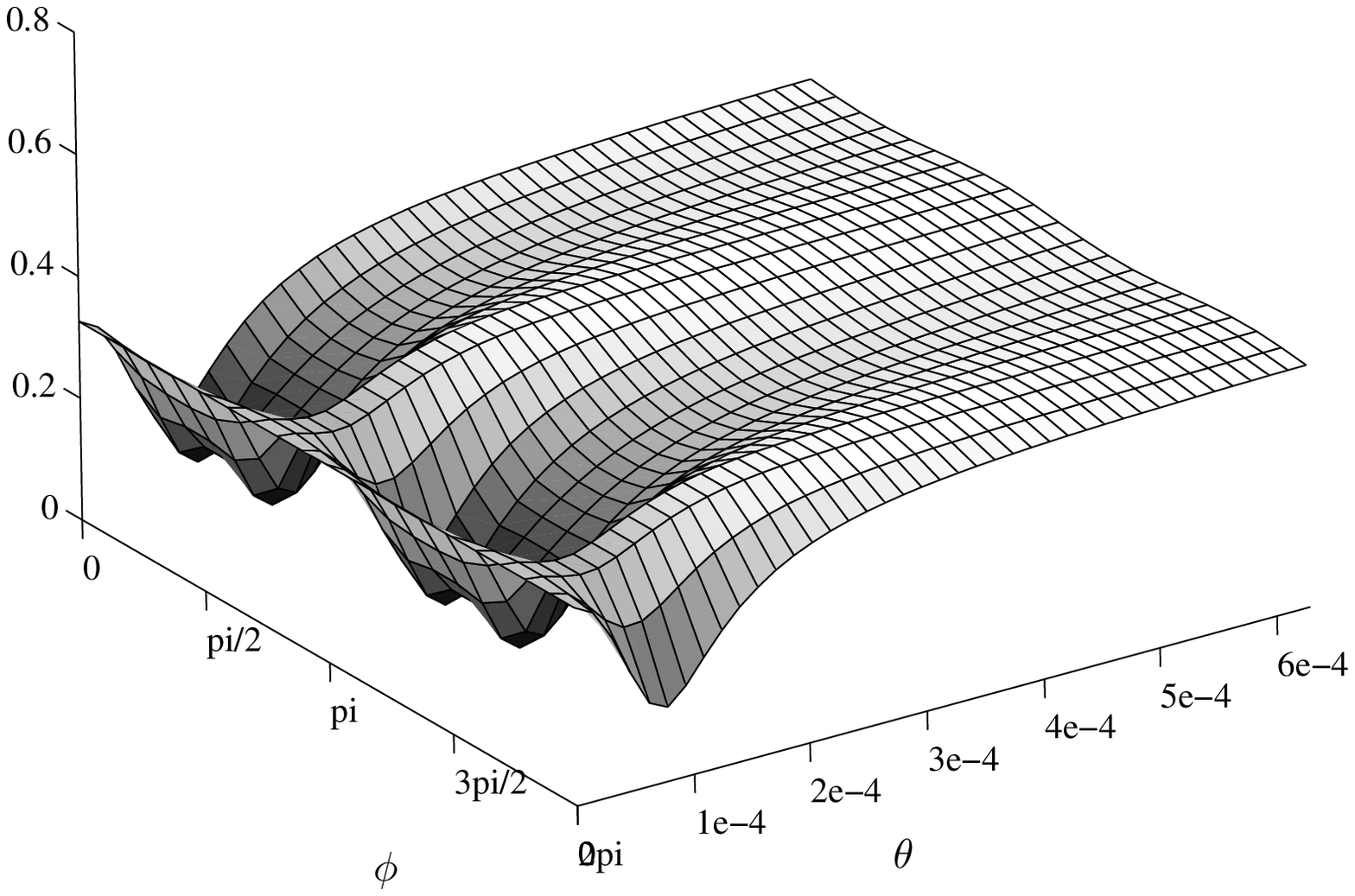}\\
\caption{\label{FIG_II_1}\footnotesize{The angular dependence of the first harmonic ($n=1$). The plots show the quantity  $\tilde{h}\cdot\frac{r}{2 \pi G\mu}\frac{\left(1-{\bf v}\cos\theta  \right)}{V^2\beta^3}$. We have set $G\mu=10^{-9}$. The right panel is a closeup on the value of $\theta$ of order $V\beta\simeq1.5\cdot10^{-4}$.}}
\end{center}
\end{figure}

\begin{figure}
\begin{center}
\includegraphics[width=10cm]{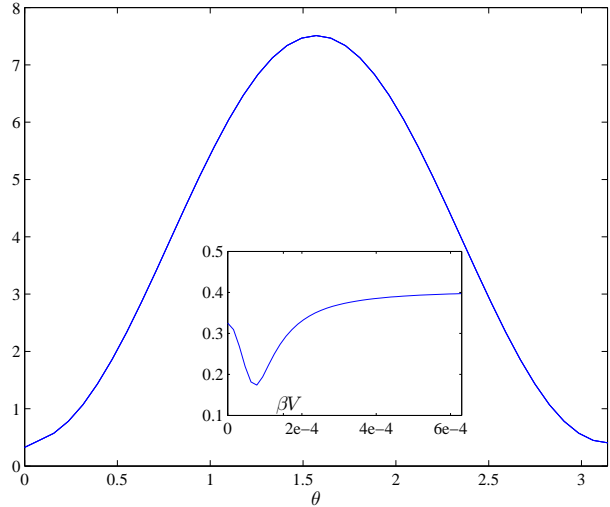}
\caption{\label{FIG_II_2}\footnotesize{Same as Figure~\ref{FIG_II_1} but after averaging over $\phi$. Note that for $V\beta < \theta < \pi-V\beta$ the curve is approximately proportional to $\sin^2\theta$. This dependence arises from the rotation and posterior conversion to TT gauge.}}
\end{center}
\end{figure}

We note that the main dependency of $\tilde{h}$ in the direction of the observation comes from the $f_1(\theta)$ factor in Eq.~(\ref{h01}) which, for small values of $\theta$, causes an enhancement by a factor $(V\beta)^{-2}$ ($\sim10^7$ for $G\mu=10^{-9}$ ), as was discussed in Section~\ref{S_PL}.  The remaining angular dependence arising from $\mathcal{I}^{\mu\nu}(\theta,\phi)$, for the first harmonic, is shown in Figures~\ref{FIG_II_1} and~\ref{FIG_II_2}.  One can observe that this gives some enhancement\footnote{This enhancement is a consequence of the rotation between the source frame and the observer frame.  This is more clearly seen by taking $\phi=0$, in which case the dominant contribution to $\tilde{h}/f_1(\theta)$ is proportional to $\sin^2(\theta)$.} for $\theta\sim\pi/2$.  However, this corresponds to the range in the parameter $\theta$ where the large Lorentz boost enhancement is lost.  Let us anticipate that in order to be able to detect the loop we need a big enhancement of the emitted GW (in other terms, we are able to `see' only loops moving nearly in our direction).  Therefore any sub-leading effect at large $\theta$ may be neglected and in the remaining of this paper we will discard this sub-leading dependence. This leads to the following form (omitting the higher harmonics) for $\tilde{h}$:
\bea
\tilde{h}&\simeq&\frac{8 \pi G\mu}{r}\frac{V^2\beta^3}{\left(1-|\vtr{v}|\cos\theta \right)}\left\vert \int_{-1}^1xe^{i\pi x^3} dx \right\vert^2\cdot0.5 \nonumber \\
&\simeq&1.3\frac{\pi}{r}\frac{G\mu V^2\beta^3}{\left(1-\cos\theta+\frac{V^2\beta^2}{6}\cos{\theta}\right)} \ ,
\label{e:h}
\eea
where the final factor of 0.5 shifts the value of the integral at $\theta=0$ to its actual minimal value (see Figure~\ref{FIG_II_2}) in order to obtain a lower estimate for the strain $\tilde{h}$.

\section{Visibility\label{S_V}}

Having computed the GW spectrum, we now want to determine whether or not current or planed GW detectors can observe the small relativistic loops under consideration and, if so, for what range of the parameters.  This will depend on the cosmology and therefore we must fix the $z$ dependence of $t$ and $r$.  To this end, we consider a flat universe with the cosmological constant contribution included in the cosmology.  We shall use the same relations (and notation) as the ones given in appendix B of~\cite{Siemens:2006vk}.

Any GW detector is characterized by its frequency window $[f_{-};f_{+}]$ as well as the minimal amplitude needed for a detection $h_{\rm det}(f)$.  For LIGO's second science run (LIGO S2) the frequency window is $[160;728.8]Hz$.  Furthermore, the sensitivity to a continuous signal from a 10-hour search can be modeled by
\be
h_{\rm det}(f) = h_{160}\left(\frac{f}{160 Hz}\right)^\alpha \ ,
\ee
where $\alpha\sim0.6$ and $h_{160}\sim10^{-22}$, see~\cite{Abbott:2006vg}.  LIGO is expected to be 10 times more sensitive and Advanced LIGO to bring another factor of 10 (see for example~\cite{Creighton:2003nm}). 
  
In order to detect a given loop there are 3 conditions which must be met: first, the frequency of the incoming GW produced by the loop has to lie within the detector window; second, its amplitude has to be above the limit $h_{\rm det}(f)$; finally, the observed lifetime of the loop, $\tau_{\rm obs}$, must be large compared to the time span of the experiment, $T_{\rm obs}$, to insure that its periodicity does not change significantly during observation, in which case it would quickly drop out of the frequency window anyway.

\begin{figure}
\begin{center}
\includegraphics[width=10cm]{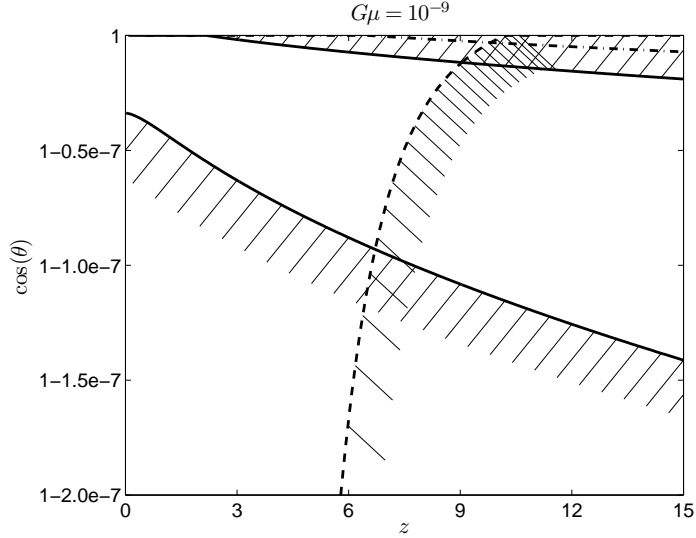}
\caption{\label{FIG_wnd}\footnotesize{Observational window for LIGO and for $G\mu=10^{-9}$.  We plot $\cos(\theta)$ as a function of the red-shift.  The solid curves represent the frequency window, the dashed curve is the sensitivity bound and the dash-dotted curve corresponds to the lifetime constraint.  Note that, for such a value of $G\mu$, we can observe loops up to $z\simeq9$ when the direction of the motion of the loop lies on a cone of aperture $\sim2\cdot10^{-4}rad\simeq0.7'$ around the line of sight. }}
\end{center}
\end{figure}

Inspection of equation~(\ref{harmonics}) shows that the frequency of the received GW depends on the angle $\theta$ subtended between the line of sight and the velocity of the loop.  Thus, the strategy is to first fix $G\mu$ and determine the range of angles under which observation is possible, for a given $z$.  We shall give the different bounds in terms of the cosine of the corresponding angle.  Then, for each value of $G\mu$, combination of the above constraints determines an observational window.  An example with $G\mu=10^{-9}$ is shown in Figure~\ref{FIG_wnd}.  We point out that, if our assumptions are correct, one can potentially observe these loops if $10^{-10}\lsim G\mu\lsim 7\cdot10^{-9}$.  Finally, the integration over the redshift can be performed to determine the expected number of detections.  The capacity of observing such loops (or placing constraints on the string tension) will then depend on their number density in the universe, and this quantity will be worked out in Section~\ref{S_LDPD}.  But before we do so, let us turn to the analysis of the several constraints mentioned above.

\subsection*{Constraint from the frequency window}
  
In order to determine the observed frequency of a GW produced by a small loop we must correct equation~(\ref{harmonics}) for the cosmological redshift, thus obtaining
\be
f_n\vert_{\rm observed}=\frac{1}{\beta(1+z)}\frac{n}{(1-\cos\theta(1-V^2\beta^2/6))} \ .
\ee  
The bounds $f_\pm$ on the observed frequency lead to corresponding constraints on the angle $\theta$ which can be expressed as $\cos\theta_- \leq \cos\theta \leq \cos\theta_+$, with
\be
\cos\theta_\pm = \frac{1-\frac{n}{f_{\pm} \beta (1+z)}}{1-\frac{1}{6}V^2\beta^2} \ .
\label{bfreq}
\ee

\subsection*{Constraint from the sensitivity}

The amplitude $\tilde{h}$ of the GW wave produced by a small cosmic string loop was obtained in~(\ref{e:h}).  Taking into account the cosmology and converting to the time domain, the strain for the lowest harmonic takes the following form:
\be
h(t,\theta) = \frac{1.3\ G\mu V^2\beta^3}{(1-\cos\theta(1-V^2\beta^2/6)) \,r(z)} 
\cos(2\pi f_1(\theta)t-\psi_0) \ ,
\ee
where $\psi_0$ is an (arbitrary) phase.  The condition to be satisfied in order to be able to detect the GW is then $h(t,\theta) \geq h_{\rm det}(f_1\vert_{\rm observed}(\theta))$.  Solving for the angle yields a lower bound, $\cos\theta \geq \cos\theta_h$, where
\be
\cos\theta_h = \frac{1-\left(\frac{1.3\ G\mu V^2\beta^{3+\alpha }(1+z)^\alpha(160Hz)^\alpha}{h_{160}\,r(z)}\right)^{1/(1-\alpha)}}{1-\frac16V^2\beta^2} \ . \label{c:s}
\ee

\subsection*{Constraint from the lifetime}

Another feature that must be taken into account is that cosmic string loops shrink with time due to the loss of energy in the form of gravitational waves themselves.  The power radiated is given by $\dot{E} = \Gamma G\mu$, with $\Gamma \sim 50$~\cite{Burden:1985md,Quashnock:1990wv,VilenkinShellard}.  Thus, the lifetime of a loop of size $2\beta$ in the FRW frame is $\tau=(2\beta)(\Gamma G\mu)^{-1}$.  Since the loops are moving with a relative velocity which makes an angle $\theta$ with the line of sight their observed lifetime, including the cosmological redshift, is
\be
\tau_{\rm obs} = (1-|\vtr{v}|\cos\theta)(1+z)\tau = (1-|\vtr{v}|\cos\theta)(1+z) \frac{2\beta}{\Gamma G\mu} \ .
\label{lifetime}
\ee
Given the extremely high velocities of the loops, one finds that for very small angles $\theta$ this apparent lifetime can be very short since it is suppressed by a factor of $\gamma^2$.  In that case, if we want to consider loops whose frequency does not change significantly during an observation time $T_{\rm obs}$ we need to guaranty that $\tau_{\rm obs} \geq T_{\rm obs}$, which translates into the constraint $\cos\theta \leq \cos\theta_L$, where
\be
\cos\theta_L = \frac{1-\frac{\Gamma G\mu T_{\rm obs}}{2\beta(1+z)}}{1-\frac{1}{6}(V\beta)^2} \ .\label{blife}
\ee
This bound has the same form as the one coming from the frequency window~(\ref{bfreq}).  The lifetime constraint can thus be recast as a constraint on the frequency window instead.  Combining equation~(\ref{bfreq}) and~(\ref{blife}) we obtain an effective maximal frequency given by $f^{\rm eff}_+=\min\left\{f_+,f_L\right\}$, with
\be
f_L = \frac{2n}{\Gamma G\mu T_{\rm obs}} \ .
\ee

The lifetime constraint has the effect of introducing a $G\mu$-dependent maximal frequency.  For the first harmonic ($n=1$) and a time span of $T_{obs}=10$ hours in the settings of LIGO, this does not affect the original frequency window for $G\mu \leq 1.5\cdot10^{-9}$.  Beyond that, the maximal frequency is given by $f_L$, which decreases as $G\mu$ is increased, until it finally closes the frequency window for $G\mu \simeq 6.9\cdot10^{-9}$.  If we want to observe or constrain higher values of $G\mu$ we have to look at higher harmonics or make a shorter run.  Both possibilities help relax the bound coming from the lifetime but on the other hand they also tighten the sensitivity constraint.

\section{Loop number density and expected number of detections\label{S_LDPD}}

We assume that the string network is in a scaling regime.  The rate at which the long string is converted into small loops can be obtained from equation~(4.24) of~\cite{Polchinski:2006ee}.  Inserting the numbers for a matter-dominated era, where the energy density in long strings is $\rho_\infty \simeq 4 \mu t^{-1}$, one finds that
\be
\left(\frac{\partial\ell_\infty}{\partial t}\right)_{\rm loops} \simeq \frac{4{\rm Vol}}{5\,t^3} \ ,
\ee
where ${\rm Vol}$ is the Hubble volume.  We obtain the number density in small loops by multiplying the above equation by the lifetime and dividing by their length as well as the volume factor.  Since the lifetime of the loop~(\ref{lifetime}) introduces a dependency on the angle of observation $\theta$, the apparent number density is given by
\be
n(\theta) = (1-|\vtr{v}|\cos\theta)n_{FRW} \equiv (1-|\vtr{v}|\cos\theta) \frac{4}{5}\frac{1}{\Gamma G\mu}\frac{1}{t^3} \ .
\ee

Now we can include the constraints from the Section~\ref{S_V}.  Only loops moving along certain directions and within certain distances emit GW which enter the frequency window with an amplitude high enough to be detected.  The range of $\theta$ for which the detector can `see' the loop is between $\cos\theta_{max}=\max\left\{-1,\min\left\{\cos\theta_+,\cos\theta_L\right\}\right\}$ and $\cos\theta_{min}=\min\left\{1,\max\left\{-1,\cos\theta_-,\cos\theta_h\right\}\right\}$. 
Therefore, averaging over the sphere gives
\be
\frac12\int_{\theta_{min}}^{\theta_{max}} n(\theta) \sin\theta \, d\theta =\underbrace{\frac12\left(\cos\theta_{max}-\cos\theta_{min}-\frac{|\vtr{v}|}{2}\cos^2\theta_{max} + \frac{|\vtr{v}|}{2}\cos^2\theta_{min}\right)}_{\zeta(\theta_{max},\,\theta_{min})} n_{FRW} \ .
\ee

Using the notation of~\cite{Siemens:2006vk}, where the cosmology is encoded in the function $h(z)$ and the comoving variables are expressed as $t=\frac{\varphi_t(z)}{H_0}$ and $r=\frac{\varphi_r(z)}{H_0}$, the number of observed loops is simply
\be
N=\int \zeta(\theta_{max},\theta_{min}) \, n_{FRW} \, dV \ ,
\ee
where the comoving volume is given by
\be
dV=\frac{4\pi}{H_0^3}\frac{\varphi^2_r(z)}{\left(1+z\right)^3h(z)} dz\ .
\ee
and $H_0$ represents the current value of the Hubble parameter.  Therefore, we obtain the following expression for the expected number of observations:
\be
N= \frac{16\pi}{5\Gamma G\mu}\int_0^\infty \frac{\varphi_r^2(z)}{\varphi_t^3(z)}\frac{ \zeta(\theta_{max},\theta_{min})}{(1+z)^3h(z)} \, dz \ .
\ee

\begin{figure}
\begin{center}
\includegraphics[width=10cm]{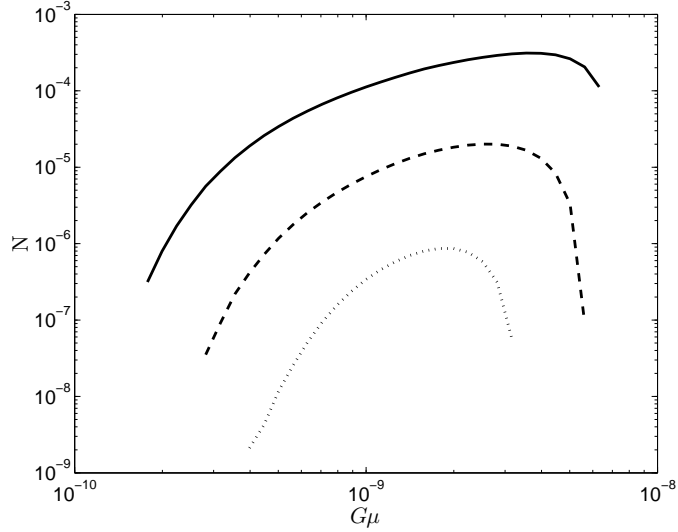}
\caption{\label{FIG_res}\footnotesize{Expected number of observations for a run of $T_{obs}=10$ hours.  The dotted, dashed and continuous curves represent LIGO S2, LIGO and Advanced LIGO, respectively.} }
\end{center}
\end{figure}

As a result, we report in Figure~\ref{FIG_res} the expected number of detections for a run of 10 hours with the different versions of LIGO.  We can understand heuristically some qualitative features of the curves.  The rise of the curves for increasing $G\mu$ comes mainly from the dependence of the GW strength on this parameter (see equation~(\ref{e:h})) which allows to observe a larger volume (equation~(\ref{c:s}) shows how the sensitivity constraint becomes less restraining with increasing $G\mu$), while the abrupt cutoff at large values of $G\mu$ has its origin in the closing of the frequency window by the upper bound from the lifetime constraint. Another effect that contributes to this fall-off is the fact that the regions allowed by the frequency window and the sensitivity bound do not overlap anymore for high $G\mu$.

Note that reducing the observation time has 2 effects: the first one is to drop the GW strength relative to the detector sensitivity since $h_{160}$ is proportional to $T_{\rm obs}^{-1/2}$~\cite{Abbott:2006vg}.  On the other hand, this allows the detection of short-lived loops at higher redshifts by relaxing the lifetime constraint and consequently we can scan larger values of $G\mu$.  However, this leaves the expected number of observations far from being of order 1 or larger.

Note also that we have only taken into account the first harmonic.  Adding higher harmonics has the effect of expanding the curves in Figure~\ref{FIG_res} to higher values of $G\mu$ but this does not cause the curves to rise.
For specific combinations of the parameters $G\mu$ and $z$ it is possible to observe more than one harmonic in the detector frequency window.  This may provide (through a specific search for direction correlated signals) better sensitivity.  However, this is certainly not enough to get an expected number of observations of order 1.
  
\section{Concluding remarks}

In conclusion, we have shown that the observation of boosted cosmic string loops is, in principle, possible for very interesting values of $G\mu$ in the range $10^{-10}-10^{-8}$.  However, the chances of such an observation during a 10 hour period are slim due to the low effective number density of such loops.  Even though a large boost factor can bring the typical frequency of the GW produced into the LIGO frequency band, this will only occur if the loop in question is moving along a trajectory very close to the line of sight ($\theta \lsim (V\beta)^2$), which effectively cuts down the apparent number density of loops with these characteristics.  An improvement can be obtained by increasing the duration of observation, thereby gaining in sensitivity of the detector.  If one keeps increasing the observation time, the apparent lifetime of the loops will eventually be reached.  We point out that, due to the production mechanism of the small loops (most of the loops that are produced around a large cusp move in coincident directions), we do not expect a Poisson distribution.  As a result, while longer times of observation lead to higher probabilities of detection, the expected number $N$ of loops observed should grow less than linearly with $T_{\rm obs}$.

We should stress that the possibility of observing GW from cosmic strings we have considered in this article concerns only the continuous emission of GW by the lower harmonics of the loops.  This is in contrast with~\cite{Damour:2004kw,Siemens:2006vk,Damour:2000wa,Damour:2001bk} which focus on the burst of GW produced by the cusps and kinks.  The results we have presented here depend both on the exact characteristics of the population of small loops (namely, their size) and also on the type of cosmic string network.  We have assumed the distribution of loop size to be sharply peaked at the gravitational radiation scale but a decaying power law would reproduce a more realistic population.  Also, cosmic superstrings can have reconnection probabilities as small as $p \sim 10^{-3}$~\cite{Jackson:2004zg} and the number density of loops is proportional to a negative power of $p$.  Numerical studies~\cite{Sakellariadou:2004wq,Avgoustidis:2005nv} suggest values between $-1$ and $-0.6$ for the exponent.  This would lead to an enhancement of the probability of detection but not quite enough to obtain $N \sim 1$.  On the other hand, our conclusions are independent of the presence or absence of high frequency features on the loops, i.e. cusps and kinks.

Note also that the discrimination between a signal from a boosted loop and other periodic signals is not difficult since the former is expected to have a rising frequency and a decreasing amplitude.  This is in contrast with both spinning stars, which are expected to spin down, and with mergers, which see their GW amplitude growing with time.  However, a complete understanding of the waveform is not possible without a description of the higher harmonics and is beyond the scope of this paper.

Finally, we have restricted our analysis to the LIGO frequency band.  For the lower frequencies of the LISA band the potential sources are loops with high $G\mu$ at very large $z$.  However, in order to get a sufficient signal-to-noise ratio we have to consider longer observation times and then the corresponding lifetime bound closes the observational window.

\section*{Acknowledgments }
  
We thank Joe Polchinski for early collaboration as well as helpful discussions and comments.  J.V.R. also thanks Xavier Siemens for useful discussions.  The work of F.D. is supported by the Swiss National Funds and by the National Science Foundation under Grant No. PHY05-51164.  J.V.R. acknowledges financial support from {\it Funda\c{c}\~ao para a Ci\^encia e a Tecnologia}, Portugal, through grant SFRH/BD/12241/2003.



\begin{thebibliography}{99}
\itemsep = 3pt

\bibitem{Polchinski:2007rg}
  J.~Polchinski and J.~V.~Rocha,
  ``Cosmic String Structure at the Gravitational Radiation Scale,''
  [arXiv:gr-qc/0702055].
 
\bibitem{Sarangi:2002yt}
  S.~Sarangi and S.~H.~H.~Tye,
  ``Cosmic string production towards the end of brane inflation,''
  Phys.\ Lett.\  B {\bf 536}, 185 (2002)
  [arXiv:hep-th/0204074].
  
\bibitem{Copeland:2003bj}
  E.~J.~Copeland, R.~C.~Myers and J.~Polchinski,
  ``Cosmic F- and D-strings,''
  JHEP {\bf 0406}, 013 (2004)
  [arXiv:hep-th/0312067].

\bibitem{Polchinski:2004ia}
  J.~Polchinski,
  ``Introduction to cosmic F- and D-strings,''
  arXiv:hep-th/0412244.
  
\bibitem{Davis:2005dd}
  A.~C.~Davis and T.~W.~B.~Kibble,
  ``Fundamental cosmic strings,''
  Contemp.\ Phys.\  {\bf 46}, 313 (2005)
  [arXiv:hep-th/0505050].
    
\bibitem{Albrecht:1989mk}
  A.~Albrecht and N.~Turok,
  ``Evolution of cosmic string networks,''
  Phys.\ Rev.\  D {\bf 40}, 973 (1989).
  
\bibitem{Bennett:1989yp}
  D.~P.~Bennett and F.~R.~Bouchet,
  ``High resolution simulations of cosmic string evolution. 1. Network evolution,''
  Phys.\ Rev.\  D {\bf 41}, 2408 (1990).
  
\bibitem{Allen:1990tv}
  B.~Allen and E.~P.~S.~Shellard,
  ``Cosmic string evolution: a numerical simulation,''
  Phys.\ Rev.\ Lett.\  {\bf 64} (1990) 119.
  
\bibitem{Hindmarsh:1994re}
  M.~B.~Hindmarsh and T.~W.~B.~Kibble,
  ``Cosmic strings,''
  Rept.\ Prog.\ Phys.\  {\bf 58}, 477 (1995)
  [arXiv:hep-ph/9411342].

\bibitem{VilenkinShellard}
	A.~Vilenkin, E.~P.~S.~Shellard,
	``Cosmic Strings and Other Topological Defects,''
 	{\it Cambridge University, Cambridge, MA, 2000}

\bibitem{Ringeval:2005kr}
  C.~Ringeval, M.~Sakellariadou and F.~Bouchet,
  ``Cosmological evolution of cosmic string loops,''
  [arXiv:astro-ph/0511646].
   
\bibitem{Vanchurin:2005pa}
  V.~Vanchurin, K.~D.~Olum and A.~Vilenkin,
  ``Scaling of cosmic string loops,''
  Phys.\ Rev.\  D {\bf 74}, 063527 (2006)
  [arXiv:gr-qc/0511159].
  
\bibitem{Olum:2006ix}
  K.~D.~Olum and V.~Vanchurin,
  ``Cosmic string loops in the expanding universe,''
  [arXiv:astro-ph/0610419].

\bibitem{Polchinski:2006ee}
  J.~Polchinski and J.~V.~Rocha,
  ``Analytic study of small scale structure on cosmic strings,''
  Phys.\ Rev.\  D {\bf 74}, 083504 (2006)
  [arXiv:hep-ph/0606205].

\bibitem{Vachaspati:1984gt}
  T.~Vachaspati and A.~Vilenkin,
  ``Gravitational Radiation From Cosmic Strings,''
  Phys.\ Rev.\  D {\bf 31}, 3052 (1985).
  
\bibitem{Damour:2004kw}
  T.~Damour and A.~Vilenkin,
  ``Gravitational radiation from cosmic (super)strings: Bursts, stochastic background, and observational windows,''
  Phys.\ Rev.\  D {\bf 71}, 063510 (2005)
  [arXiv:hep-th/0410222].
  
\bibitem{Siemens:2006vk}
  X.~Siemens, J.~Creighton, I.~Maor, S.~Ray Majumder, K.~Cannon and J.~Read,
  ``Gravitational wave bursts from cosmic (super)strings: Quantitative analysis and constraints,''
  Phys.\ Rev.\  D {\bf 73}, 105001 (2006)
  [arXiv:gr-qc/0603115].
  
\bibitem{Hogan:2006we}
  C.~J.~Hogan,
  ``Gravitational waves from light cosmic strings: Backgrounds and bursts with large loops,''
  Phys.\ Rev.\  D {\bf 74}, 043526 (2006)
  [arXiv:astro-ph/0605567].

\bibitem{Economou:1991bc}
  A.~Economou, D.~Harari and M.~Sakellariadou,
  ``Gravitational effects of traveling waves along global cosmic strings,''
  Phys.\ Rev.\  D {\bf 45}, 433 (1992).
  
\bibitem{Battye:1997ji}
  R.~A.~Battye, R.~R.~Caldwell and E.~P.~S.~Shellard,
  ``Gravitational waves from cosmic strings,''
  [arXiv:astro-ph/9706013].

\bibitem{Siemens:2006yp}
  X.~Siemens, V.~Mandic and J.~Creighton,
  ``Gravitational wave stochastic background from cosmic (super)strings,''
  [arXiv:astro-ph/0610920].
  
\bibitem{Damour:2000wa}
  T.~Damour and A.~Vilenkin,
  ``Gravitational wave bursts from cosmic strings,''
  Phys.\ Rev.\ Lett.\  {\bf 85}, 3761 (2000)
  [arXiv:gr-qc/0004075].
  
\bibitem{Damour:2001bk}
  T.~Damour and A.~Vilenkin,
  ``Gravitational wave bursts from cusps and kinks on cosmic strings,''
  Phys.\ Rev.\  D {\bf 64}, 064008 (2001)
  [arXiv:gr-qc/0104026].
  
\bibitem{Thompson:1988yj}
  A.~C.~Thompson,
  ``Dynamics of cosmic string,''
  Phys.\ Rev.\  D {\bf 37}, 283 (1988).
  
\bibitem{Quashnock:1990wv}
  J.~M.~Quashnock and D.~N.~Spergel,
  ``Gravitational selfinteractions of cosmic strings,'' 
   Phys.\ Rev.\  D {\bf 42}, 2505 (1990).
  
\bibitem{Siemens:2001dx}
  X.~Siemens and K.~D.~Olum,
  ``Gravitational radiation and the small-scale structure of cosmic  strings,''
  Nucl.\ Phys.\  B {\bf 611}, 125 (2001)
  [Erratum-ibid.\  B {\bf 645}, 367 (2002)]
  [arXiv:gr-qc/0104085].
  
\bibitem{Siemens:2003ra}
  X.~Siemens and K.~D.~Olum,
  ``Cosmic string cusps with small-scale structure: Their forms and gravitational waveforms,''
  Phys.\ Rev.\  D {\bf 68}, 085017 (2003)
  [arXiv:gr-qc/0307113].
  
\bibitem{Chialva:2006ak}
  D.~Chialva and T.~Damour,
  ``Quantum effects in gravitational wave signals from cuspy superstrings,''
  JCAP {\bf 0608}, 003 (2006)
  [arXiv:hep-th/0606226].

\bibitem{Garfinkle:1987yw}
  D.~Garfinkle and T.~Vachaspati,
  ``Radiation from kinky, cuspless cosmic loops,''
  Phys.\ Rev.\  D {\bf 36}, 2229 (1987).

\bibitem{Pogosian:2006hg}
  L.~Pogosian, I.~Wasserman and M.~Wyman,
  ``On vector mode contribution to CMB temperature and polarization from  local
  strings,''
  arXiv:astro-ph/0604141.

\bibitem{Jeong:2006pi}
  E.~Jeong and G.~F.~Smoot,
  ``Validity of cosmic string pattern search with cosmic microwave
  background,''
  arXiv:astro-ph/0612706.

\bibitem{Misner:1974qy}
  C.~W.~Misner, K.~S.~Thorne and J.~A.~Wheeler,
  ``Gravitation,''
	{\it  W. H. Freeman, San Francisco, 1973}

\bibitem{Weinberg}
	S.~Weinberg,
	``Gravitation and Cosmology: Principles and Applications of the General Theory of Relativity,''
	{\it  John Wiley \& Sons Inc, New York, 1972 }

\bibitem{Abbott:2006vg}
  B.~Abbott {\it et al.}  [LIGO Scientific Collaboration],
  ``Coherent searches for periodic gravitational waves from unknown isolated
  sources and Scorpius X-1: Results from the second LIGO science run,''
  [arXiv:gr-qc/0605028].

\bibitem{Creighton:2003nm}
  T.~Creighton,
  ``Advanced Ligo: Sources And Astrophysics,''
  Class.\ Quant.\ Grav.\  {\bf 20}, S853 (2003).

\bibitem{Burden:1985md}
  C.~J.~Burden,
  ``Gravitational Radiation From A Particular Class Of Cosmic Strings,''
  Phys.\ Lett.\  B {\bf 164}, 277 (1985).

\bibitem{Jackson:2004zg}
  M.~G.~Jackson, N.~T.~Jones and J.~Polchinski,
  ``Collisions of cosmic F- and D-strings,''
  JHEP {\bf 0510}, 013 (2005)
  [arXiv:hep-th/0405229].

\bibitem{Sakellariadou:2004wq}
  M.~Sakellariadou,
  ``A note on the evolution of cosmic string / superstring networks,''
  JCAP {\bf 0504}, 003 (2005)
  [arXiv:hep-th/0410234].

\bibitem{Avgoustidis:2005nv}
  A.~Avgoustidis and E.~P.~S.~Shellard,
  ``Effect of reconnection probability on cosmic (super)string network
  density,''
  Phys.\ Rev.\  D {\bf 73}, 041301 (2006)
  [arXiv:astro-ph/0512582].

\end{thebibliography}
\end{document}